# Electrically switchable Rashba-type Dzyaloshinskii-Moriya interaction and skyrmion in two-dimensional magnetoelectric multiferroics

Jinghua Liang[1], Qirui Cui[1], Hongxin Yang[1,2*]

[1] *Ningbo Institute of Materials Technology and Engineering, Chinese Academy of Sciences, Ningbo 315201, China*

[2] *Center of Materials Science and Optoelectronics Engineering, University of Chinese Academy of Sciences, Beijing 100049, China*

Realizing topological magnetism and its electric control are intensive topics in the spintronics due to their promising applications in information storage and logic technologies. Here we unveil that both can be achieved in the two-dimensional (2D) magnetoelectric multiferroics. Using first-principles calculations, we show that strong Dzyaloshinskii-Moriya interaction (DMI), which is the key ingredient for the formation of exotic chiral magnetism, could be induced in 2D multiferroics with vertical electric polarization via Rashba effect. We verify that such significant DMI can promote sub-10 nm skyrmion in 2D multiferroics with perpendicular magnetic anisotropy such as CrN monolayer. In addition, the presence of both magnetization and electric polarization in 2D multiferroics provides us unique opportunity for the effective electric control of both strength and chirality of DMI and thereby the topological magnetism. As an example, we introduce the four multiferroic skyrmions with different chirality and polarity that can be manipulated by external field.

*Introduction.*―Topological chiral spin textures, e.g., spin spirals [1] and skyrmions [2, 3, 4], are of particular interest from both fundamental and applied points of view. A crucial ingredient for the generation of exotic topological magnetism is the antisymmetric Dzyaloshinskii-Moriya interaction (DMI) [5, 6] that results from spin-orbit coupling (SOC) in magnets with inversion asymmetry, favoring noncollinear spin configurations. The induction and control of DMI are essential steps towards spintronics applications of topological magnetism.

Recently, with the breakthrough in experimental realization [7, 8, 9, 10, 11], two-dimensional (2D) ferromagnetic thin films have attracted a lot of interest. However, the

---

absence of DMI in experimentally reported 2D magnets so far due to the symmetry constraint has hindered their application in skyrmion-based spintronics. In a very recent work [12], we proposed that 2D magnets with Janus structure can acquire very large DMI through the Fert-Levy mechanism, where the heavy ligands act as SOC active sites for electron hopping, yet it is unclear whether strong DMI can be induced in 2D materials via other mechanisms. Moreover, the electric control of topological magnetism in 2D magnets remains largely unexplored. We note that 2D magnetoelectric multiferroics with vertical electric polarization (hereafter referred to as 2D multiferroics unless stated otherwise) may be excellent candidates for the realization of Rashba effect induced DMI [13, 14, 15], since their structure inversion symmetry is naturally broken and intrinsic Rashba SOC additionally exists due to the electrostatic potential difference generated by the vertical electric dipole. In addition, with the simultaneous presence of magnetization and electric polarization in these multiferroics, both the strength and chirality of DMI and thereby the topological magnetism could be effectively controlled by electric field.

In this Letter, we perform first-principles calculations and micromagnetic simulations to demonstrate that 2D multiferroics, such as the recently predicted monolayers of CrN [16], $CuMP_2X_6$ (M=Cr, V; X=S, Se) [17] and vacancy-doped $CrI_3$ [18], naturally host Rashba-type DMI that can facilitate the emergence of topological magnetism. Furthermore, we show how the induced DMI and skyrmion can be controlled by an electric field.

*Proposal for electric control of DMI and topological magnetism in 2D multiferroics.*—We note that the two stable states with opposite spontaneous polarization in the double-well energy landscape can be switched reversibly by an out-of-plane electric field (see Fig. 1(a)). Concurrently, the chirality of their associated DMI is also reversed since those two states are connected by the horizontal reflection operation. Importantly, this mechanism can be utilized to manipulate the topological magnetic structures such as skyrmions. A skyrmion has the degrees of freedom of chirality, which indicates the spin swirling direction and is imposed by the chirality of DMI, and polarity, which specifies the direction of magnetization in the skyrmion core [19]. With the different combinations of directions of magnetization **m** in skyrmion core and directions of electric polarization **P**, we can realize four degenerate inter-switchable skyrmionic states with different chiralities and polarities in a single 2D multiferroics as schematically shown in Fig. 1(b). We have introduced notations A+, A−, C+ and C− as well as the corresponding icons of rotating circular arrow to distinguish the four skyrmions, where the characters A and C correspond to anticlockwise and clockwise chirality and the subscripts "+" and "−" hint the upward and downward polarity, respectively. The manipulation of inter-switchable multiferroic skyrmions can inspire new concept in the design of high-density skyrmion-based memory and logic device. In the followings, we first take CrN monolayer as a representative example to demonstrate the existence of Rashba-type DMI and skyrmions and their electric control, and then extend the discussions to other thin films.

*Computational methods.*—In the framework of micromagnetic model, the energy functional for a thin film reads

$$E[\mathbf{m}] = \int [A(\nabla \mathbf{m})^2 + D[m_z \nabla \cdot \mathbf{m} - (\mathbf{m} \cdot \nabla)m_z] - K m_z^2] d\mathbf{r}, \qquad (1)$$

where $\mathbf{m}$ is the unit vector along the direction of magnetization and $m_z$ is its *z* component. The integration in Eq. (1) is over the whole thin film, and $E[\mathbf{m}]$ is usually given in the unit of meV/formula unit (f.u.) or J/m³. $A$ is the exchange stiffness, $D$ is the DMI parameter, $K = K_\text{u} - M_s \mu_0^2/2$ is the effective anisotropy corrected by the demagnetization effect, where $K_\text{u}$ is the magnetocrystalline anisotropy, $M_s$ is the saturation magnetization and $\mu_0$ is the vacuum permeability. Here we adopt the sign convention that $D > 0$ ($D < 0$) implies anticlockwise (clockwise) DMI, and $K > 0$ ($K < 0$) indicates perpendicular (in-plane) anisotropy.

The magnetocrystalline anisotropy $K_\text{u}$ can be directly obtained by calculating the total energy difference between two magnetic states with collinear in-plane and out-of-plane magnetization directions. To calculate $D$ and $A$, we study the energy of an noncollinear homogeneous flat spin spiral $\mathbf{m} = (\sin(\boldsymbol{q} \cdot \boldsymbol{r}), 0, \cos(\boldsymbol{q} \cdot \boldsymbol{r}))$ with $\boldsymbol{q}$ being the spiral vector. Note that in the calculations, we have always considered the spin spirals propagating along the nearest-neighbor magnetic atoms, which are expected to give rise to the largest DMI energy. For small $\boldsymbol{q} = (q, 0, 0)$, Eq. (1) simplifies to

$$E[q] = Aq^2 + Dq - K/2. \qquad (2)$$

We can identify the DMI energy as $\Delta E_{DM}[q] = (E[q] - E[-q])/2$. From Eq. (2), we have the relationships $\Delta E_{DM}[q] \propto Dq$ and $(E[q] - \Delta E_{DM}[q]) \propto Aq^2$. Thus, once $E[q]$ is determined, we can simultaneously get $D$ and $A$ by the fitting to $\Delta E_{DM}[q]$ and $(E[q] - \Delta E_{DM}[q])$ for various $q$. An efficient strategy to compute $E[q]$ is to employ the generalized Bloch theorem and treat SOC within the first-order perturbation theory [20, 21]. In order to include the effect of SOC in a self-consistent way, we have followed Ref. [21] to replace the full SOC operator $H_{soc}$ by its component along the rotation axis of spiral, which is $H_{soc}^y$ in our setup, in the spin spiral calculations. With such approach, which we will refer to as qSO method, we can determine $D$ and $A$ from the self-consistently calculated spin spiral energy $E[q]$ on an equal footing.

It is convenient to introduce the dimensionless parameter

$$\kappa = \left(\frac{4}{\pi}\right)^2 \frac{AK}{D^2}. \qquad (3)$$

For $\kappa < 1$, magnetic structure of the system exhibits spin spirals as ground state, whereas for $\kappa > 1$, isolated metastable skyrmion could be formed in the background of ferromagnetic ground state [22, 23, 24].

All our first-principles calculations are performed with Vienna *ab-initio* simulation package (VASP) [25]. We have implemented the qSO method in conjunction with VASP. The exchange correlation effects are treated by the generalized gradient approximation (GGA) in the Perdew-Burke-Ernzerhof (PBE) form [26]. We have considered three types of 2D multiferroics, i.e., CrN, $CuMP_2X_6$ (M=Cr, V; X=S, Se), and vacancy-doped $CrI_3$ ($Cr_8I_{23}$, i.e., $2 \times 2$ $CrI_3$ supercell with one I vacancy). For each type of thin films, we set the energy cutoff for plane wave expansion to be 520 eV, 450 eV and 500 eV, and the Γ-centered ***k***-point mesh for the Brillouin zone integration to be 24×24×1, 16×16×1 and 6×6×1, respectively. All the structures are fully relaxed until the force acting on each atom is less than 0.001 eV/Å. The dipole moment correction is considered in all calculations.

For the micromagnetic simulation, we have used the Object Oriented MicroMagnetic Framework (OOMMF) [27], which solves the spin dynamics based on the Landau-Lifshitz-Gilbert equation [28].

*Rashba-type DMI in CrN monolayer.*—The CrN monolayer with hexagonal structure (see the top and side views of structure in Figs. 2(a) and 2(b)) is a recently predicted 2D metallic multiferroics [16]. The relaxed CrN monolayer has a lattice constant $a$ of 3.16 Å and a small buckling distance $h$ of 0.21 Å, which results in an out-of-plane electric polarization of 6.68 pC/m (indicated by the green arrow in Fig. 2(b)). Consistent with previous researches [16, 29], our calculations show that the buckled ferroelectric CrN monolayer has a lower energy of 9.08 meV/f.u. than the planar paraelectric one, which indicates that the spin-phonon coupling is responsible for its vertical ferroelectricity. The calculated saturation magnetization $M_s$ is 2.40 μ$_B$/f.u. and perpendicular magnetocrystalline anisotropy $K_u$ is 0.22 meV/f.u.

The structure inversion asymmetry induced by the vertical ferroelectric displacement could lead to the DMI between Cr atoms. To demonstrate such effect, we calculate the spin spiral energy $E[q]$ in the interval of $q$ from $-0.1\frac{2\pi}{a}$ to $0.1\frac{2\pi}{a}$ (see Fig. 2(c)). Clearly, we can see that when SOC is neglected (black points in upper part of Fig. 2(c)), spin spiral with $q$ and $-q$ is degenerate, and the ground state is the ferromagnetic state ($q = 0$). Once SOC is included, as expected, $E[q]$ (red points in upper part of Fig. 2(c)) becomes asymmetric, where the anticlockwise rotating spin spiral (see the inset in Fig. 4(b) for the illustration of spin chirality) is more favorable, due to the effect of DMI. From the zoom-in plot of $E[q]$ around $q = 0$ (inset in upper part of Fig. 2(c)), we can find that the lowest $E[q]$ locates at $q = -0.006\frac{2\pi}{a}$, corresponding to spiral period length of $\lambda = \frac{2\pi}{|q|} = 52.67$ nm. With the calculated DMI energy $\Delta E_{DM}[q] = (E[q] - E[-q])/2$ (red points in lower part of Fig. 2(c)), which shows good linear dependence on $q$, the DMI parameter $D$ and spin stiffness $A$ can be determined to be 3.74 meV Å/f.u. and 142.89 meV Å$^2$/f.u., respectively. Here the positive sign of $D$ indicates that CrN monolayer prefers anticlockwise DMI when its electric polarization **P** points along the +z direction. From the calculated magnetic parameters, we can deduce the dimensionless parameter $\kappa$ to be only 1.68, which is near the critical value ($\kappa = 1$) for the transition to spiral state so that isolated metastable skyrmion can be stabilized in CrN monolayer. We will return to this point in the latter sections.

Now, we explore the microscopic origin of the DMI in CrN monolayer. It is helpful use the atom-resolved $\Delta E_{DM}[q]$ to examine the energy source of DMI, as shown in Fig. 3(a) for $q = -0.006\frac{2\pi}{a}$. One can clearly see that $\Delta E_{DM}$ is dominated by Cr atoms, but have negligible contribution from N atoms. This feature is in great contrast to the DMI energy distribution in ferromagnet/heavy metal multilayers [30, 31] and our previous studied Janus manganese dichalcogenides [12], where the DMI energy is contributed mostly by the heavy nonmagnetic atoms that is typical for the DMI generated via the Fert-Levy mechanism. However, we can identify that it is representative for the Rashba-type DMI as in the graphene/Co heterostructure [32]. To confirm that the DMI is mediated by the conduction electrons with Rashba SOC, we plot the electronic band structures of CrN monolayer with magnetization pointing along $[1\bar{1}0]$ and $[\bar{1}10]$ directions (see Fig. 3(b)). We can see that the band structures of CrN monolayer show characteristic Rashba-type $\boldsymbol{k}$ dependent splitting. For Rashba-type DMI, the DMI parameter can be estimated as $D_R = \left(\frac{4m^*\alpha_R}{\hbar^2}\right)A$ [13], where $m^*$ is the effective mass of electron, $\alpha_R$ is the Rashba coefficient, and $A = 142.89$ meV Å$^2$/f.u. is the calculated spin stiffness. Here we focus on the electronic states that are around the Γ point and close to the Fermi level. From the zoom-in plot in Fig. 3(c), $\alpha_R = 18.0$ meV Å and $m^* = 1.94\ m_e$ with $m_e$ being the rest mass of electron can be extracted. One can then deduce that $D_R = 2.66$ meV Å/f.u., which is only slightly smaller than our previous value ($D = 3.74$ meV Å/f.u.) calculated from the qSO method. All together, we can infer that the DMI in CrN monolayer originates from Rashba effect.

*Electric control of magnetic parameters in CrN monolayer.*—Here we show that both the structural and magnetic properties of CrN monolayer can be effectively controlled by the out-of-plane electric field. Fig. 4(a) presents the variation of buckling distance *h* when the out-of-plane electric field $\varepsilon$ sweeps from 0.5 V/Å to −0.5 V/Å (black points), and vice versa (red points). From the hysteresis loop of buckling distance *h*, we can see that as $\varepsilon$ changes from 0.5 V/Å to −0.5 V/Å, *h* first increases slightly when $\varepsilon \geq 0.40$ V/Å, then deceases monotonously from 0.23 Å to 0.17 Å before the electric polarization switching occurring at $\varepsilon = -0.40$ V/Å. The variations of *h* reflect the ferroelectric displacement controlled by the external field.

Figs. 4(b)~4(d) summarize the resulting electric field dependent magnetic parameters [33] including DMI parameter $D$, spin stiffness $A$ and magnetocrystalline anisotropy $K_u$. Notably, from Fig. 4(b), we can see that the variation trend of $D$ almost follows that of *h*.

Within the field strength range of $|\varepsilon| < 0.50$ V/Å, the value of $D$ can be tuned between 3.98 meV Å/f.u. and 3.19 meV Å/f.u. with the chirality (indicated schematically in the inset of Fig. 4(b)) determined by the direction of electric polarization **P**. Our calculations thus directly validate the previous proposed electric control of both strength and chirality of DMI (see Fig. 1(a)). We also find that $A$ (Fig. 4(c)) and $K_u$ (Fig. 4(d)) generally show opposite electric field dependence, i.e., as $\varepsilon$ sweeps from 0.4 V/Å to $-0.35$ V/Å, $A$ almost decreases linearly from 172.70 meV Å$^2$/f.u. to 117.90 meV Å$^2$/f.u., while $K_u$ is enhanced from 0.19 meV/f.u. to 0.25 meV/f.u. The saturation magnetization $M_s$ of CrN monolayer is inert to the applied electric field and remains to be about 2.40 μ$_B$/f.u. in our calculations.

*Realization of skyrmion in 2D multiferroics.*—With all the magnetic parameters calculated, we can study the stability of skyrmion in CrN monolayer. As we have mentioned before, without electric field, the deduced $\kappa$ is only 1.68, which indicates the existence of isolated metastable skyrmion. To confirm this point, we perform micromagnetic simulation by relaxing a trial skyrmion spin configuration on a CrN nanodisk with radius of 30 nm. After the relaxation, we find that a metastable skyrmion with higher energy of 0.30 meV than the ferromagnetic ground state can be stabilized (see Fig. 5). The radius $R$ (defined as the radius of $m_z = 0$ contour) of relaxed skyrmion is only 4.8 nm, which is very close to $\bar{R} = 4.5$ nm predicted by the theoretical formula [34]

$$\bar{R} = \pi D \sqrt{\frac{A}{16AK^2 - \pi^2 D^2 K}}. \qquad (4)$$

For $|\varepsilon| < 0.50$ V/Å, $\kappa$ varies between 1.44 and 2.67, and the relaxed radius $R$ changes between 1.9 nm and 6.9 nm. Here we should remind that with the electric polarization switching, the chirality of skyrmion is also switched as that of DMI. The above proposed four inter-switchable multiferrroic skyrmions with tunable size (see Fig. 1(b)) can thus be realized in CrN monolayer.

To generalize the proposals for Rashba-type DMI and topological magnetism, we have considered other predicted 2D multiferroics including CuMP$_2$X$_6$ (M=Cr, V; X=S, Se) [17], and vacancy-doped CrI$_3$ (Cr$_8$I$_{23}$) [18]. Table I summarizes their calculated structural and magnetic parameters, which are also electrically switchable as for CrN monolayer. From the calculation results, we can find that they all have finite DMI, and the sizeable DMI in CuVP$_2$Se$_6$ and CuCrP$_2$Se$_6$ can even lead to a relative small $\kappa$ of ~4.0, which is similar to those in Pd/Co/Pt ($\kappa$ = 5.4) [24] and Fe/W(110) ($\kappa$ = 4.8) [35] that are predicted to host chiral magnetic solitons [24, 36]. Considering that the predicted radius $\bar{R}$ of skyrmion in CuVP$_2$Se$_6$ and CuCrP$_2$Se$_6$ is only about twice their lattice constant $a$, the classical description of $E[\mathbf{m}]$ in Eq. (1) may fail, and quantum effects should be taken into account [37], which deserves further study in the future.

**Table I.** The calculated lattice constant $a$, film thickness $t$, saturation magnetization $M_s$, DMI parameter $D$, exchange stiffness $A$, magnetocrystalline anisotropy $K_u$ and dimensionless parameter $\kappa$ for 2D multiferroic CuMP$_2$X$_6$ (M=Cr, V; X=S, Se), and vacancy-doped CrI$_3$ (Cr$_8$I$_{23}$). The radius $\bar{R}$ of skyrmion estimated by Eq. (4) is given for CuVP$_2$Se$_6$ and CuCrP$_2$Se$_6$. The sign of DMI is always determined with respect to the structure with polarization oriented along +$z$ direction.

| Structure | $a$ (Å) | $t$ (nm) | $M_s$ (μ$_B$/f.u.) | $D$ (meV Å/f.u.) | $A$ (meV Å$^2$/f.u.) | $K_u$ (meV/f.u.) | $\kappa$ | $\bar{R}$ (nm) |
|---|---|---|---|---|---|---|---|---|
| CuVP$_2$S$_6$ | 6.02 | 0.33 | 1.74 | -0.07 | 90.96 | -0.001 | -310.23 | — |
| CuVP$_2$Se$_6$ | 6.38 | 0.33 | 1.75 | 1.70 | 44.70 | 0.15 | 3.46 | 1.2 |
| CuCrP$_2$S$_6$ | 5.98 | 0.32 | 2.76 | -0.47 | 119.64 | -0.05 | -66.85 | — |
| CuCrP$_2$Se$_6$ | 6.35 | 0.34 | 2.77 | -2.50 | 111.96 | 0.17 | 4.18 | 1.6 |
| Cr$_8$I$_{23}$ | 14.01 | 0.32 | 23.27 | 10.71 | 989.41 | 3.95 | 49.69 | — |

*Conclusion.*—We have performed first-principles calculations and micromagnetic simulations to demonstrate the existence of Rashba-type DMI and skyrmion in 2D multiferroics. Due to the intrinsic coupling of magnetism and electricity in 2D multiferroics, both the strength and chirality of DMI, and thereby the size and chirality of skyrmion, can be effectively controlled by electric field. Although we have focused on 2D multiferroics with vertical polarization, our proposal for the electric control of DMI and topological magnetism can be easily extended to those with in-plane polarization [38, 39, 40], in which the DMI can be generated through other mechanism rather than the Rashba effect. In summary, our work can open up new route to the induction and electric control of topological magnetism in the emerging 2D spintronics.

*Acknowledgment.*—This work was supported by National Natural Science Foundation of China (11874059), the Key Research Program of Frontier Sciences, CAS, Grant NO. ZDBS-LY-7021, and Zhejiang Province Natural Science Foundation of China (LR19A040002).

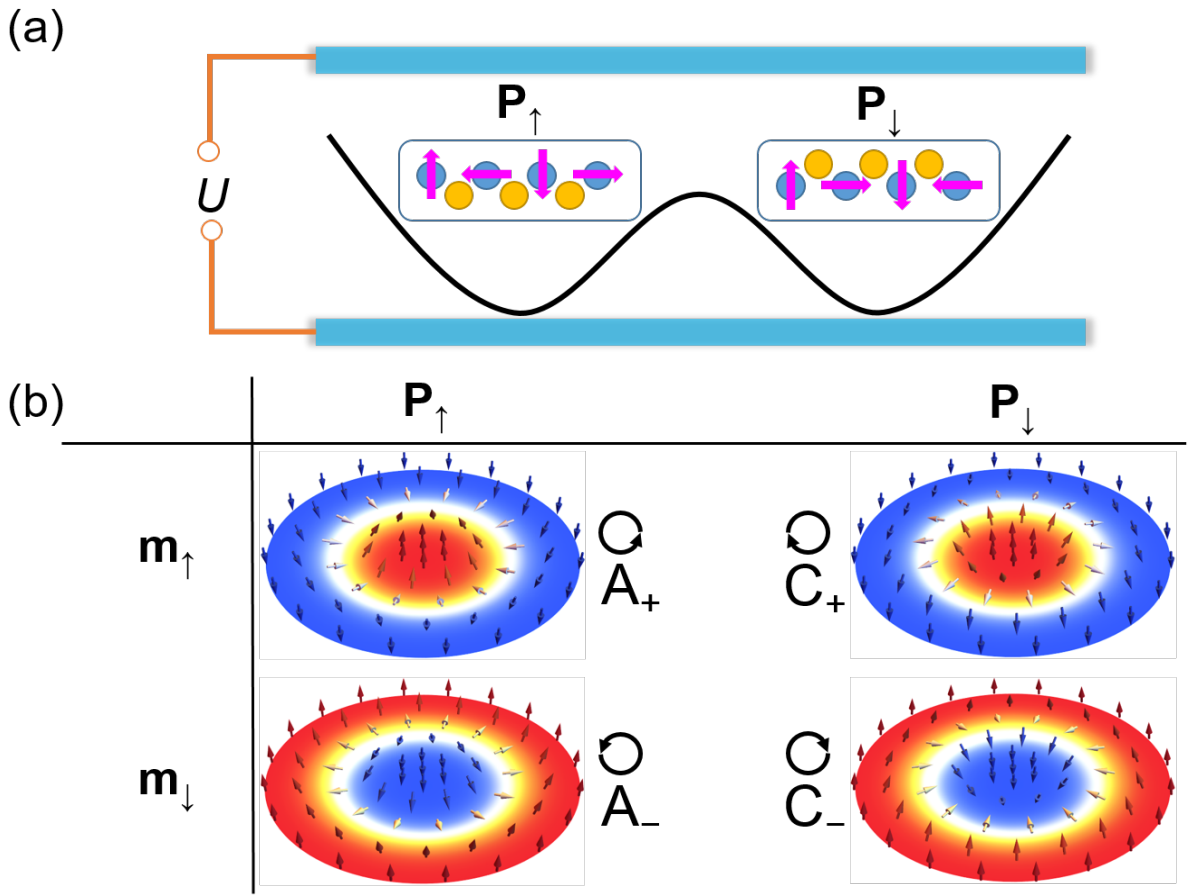

**Figure 1.** Concept of the switchable Rashba-type DMI and multiferroic skyrmions. (a) Two stable states with opposite polarization **P** in the double-well energy landscape of 2D multiferroics. With the switching of **P** by an out-of-plane electric field, the chirality (indicated by the magenta arrows) of DMI induced by Rashba effect is also reversed. (b) The four inter-switchable multiferroic skyrmions with different chirality and polarity. The arrow ↑(↓) indicates that the direction of magnetization **m** in skyrmion core or the direction of electric polarization **P** is oriented along $+z(-z)$ direction. Notations $A_+$, $A_-$, $C_+$ and $C_-$ as well as the corresponding icons of rotating circular arrow can be used to distinguish the four skyrmions.

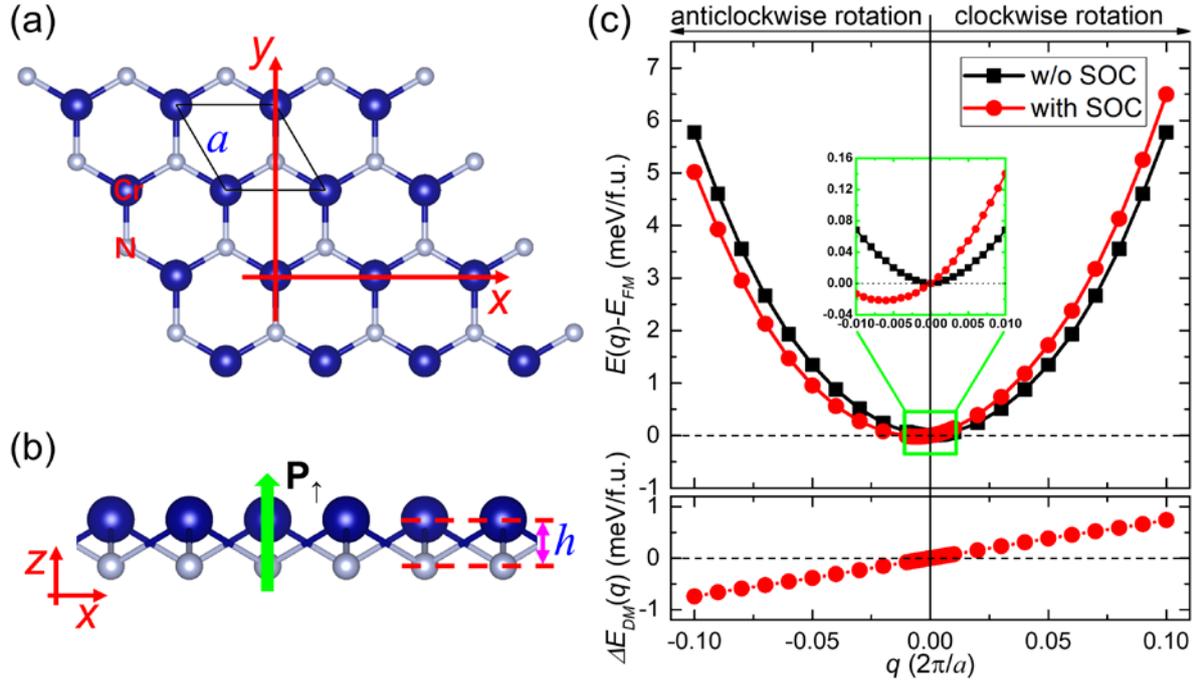

**Figure 2.** (a) Top and (b) side views of crystal structure of 2D multiferroic CrN monolayer with lattice constant $a$ and buckling distance $h$. The back lines in (a) indicate the primitive unit cell used in the calculations. (c) Spin spiral energy $E[q]$ (upper panel) and DMI energy $\Delta E_{DM}[q]$ (lower panel) calculated as a function of spiral vector length $q$. $E[q]$ is given with respect to the ferromagnetic state at $q = 0$. Black and red points are calculated without (w/o) and with SOC, respectively. Inset in the upper part of (c) is the zoom-in plot of $E[q]$ around $|q| < 0.01 \frac{2\pi}{a}$.

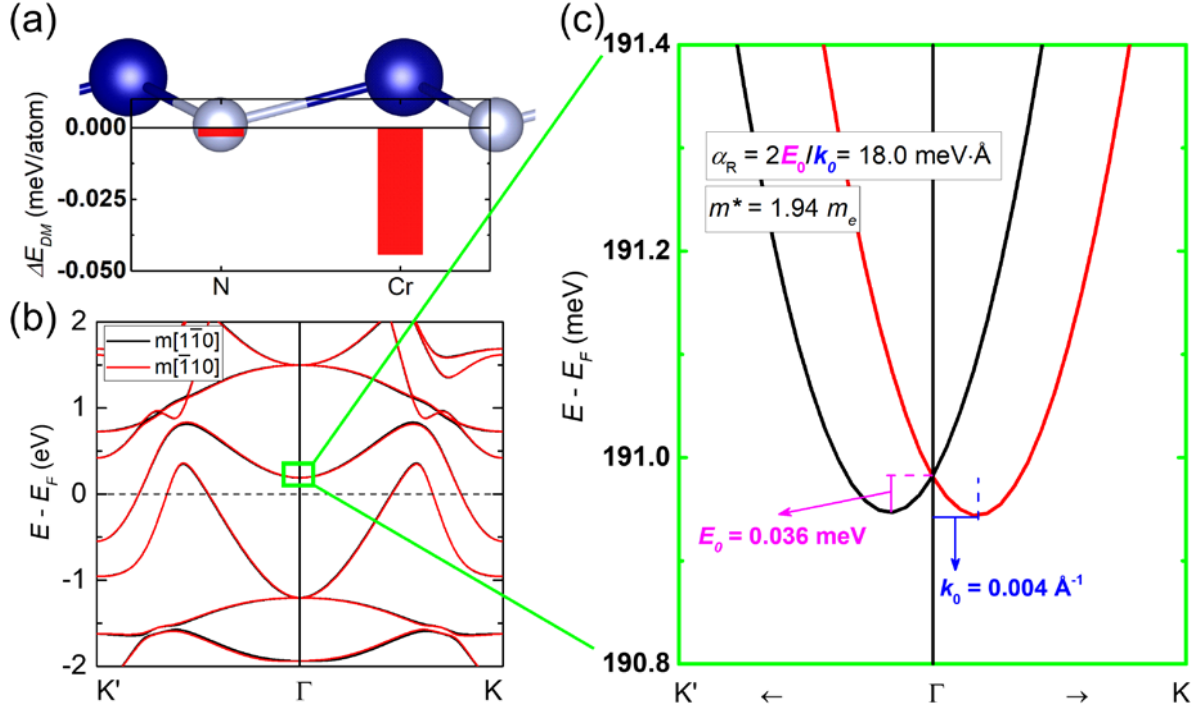

**Figure 3.** (a) Atom-resolved DMI energy $\Delta E_{DM}[q]$ at $q = -0.006\frac{2\pi}{a}$, where the minimum of $E[q]$ locates. (b) Band structures of CrN monolayer with magnetization with the magnetization direction along $[1\bar{1}0]$ (black lines) and $[\bar{1}10]$ (red lines). (c) Zoom-in plot of band structures for states that are around the $\Gamma$ point and close to the Fermi level. For these states, we can extract that the Rashba coefficient $\alpha_R = 2E_0/k_0 = 18.0$ meV Å with $E_0$ being the Rashba splitting at the wave vector $k_0$ and the effective mass $m^* = 1.94\ m_e$ with $m_e$ being the rest mass of electron.

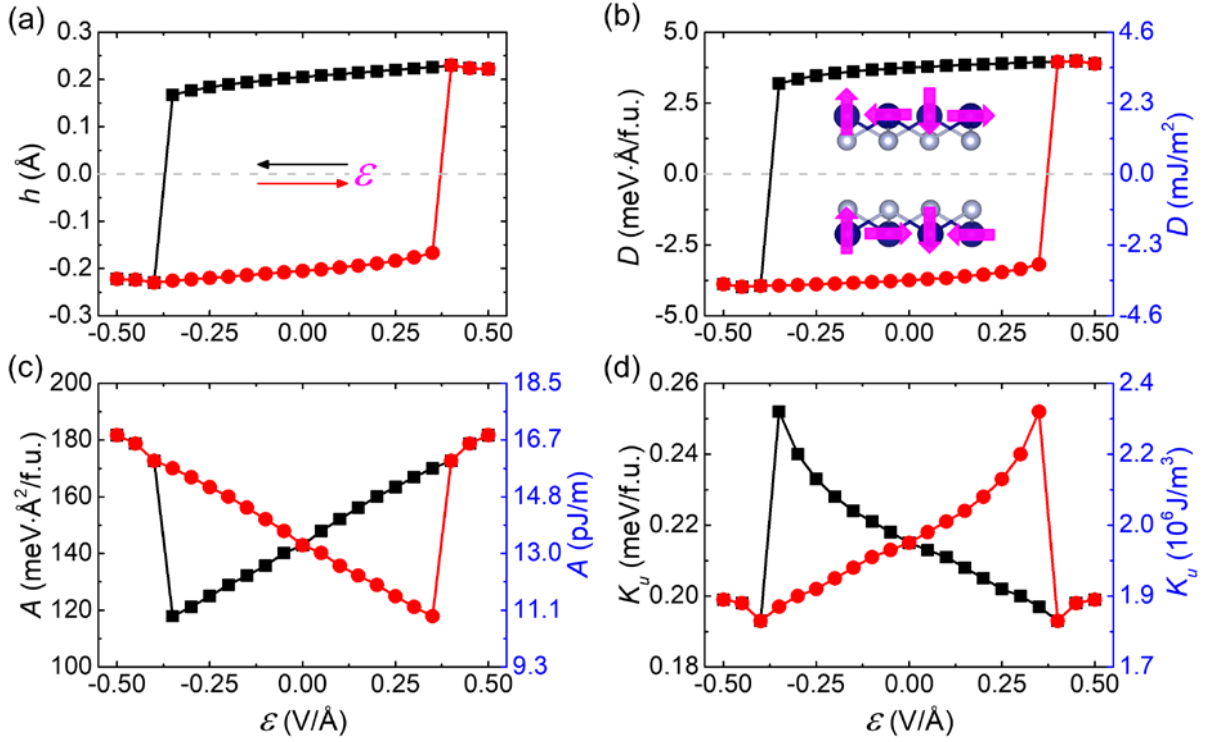

**Figure 4.** The calculated (a) buckling distance $h$, (b) DMI parameter $D$, (c) spin stiffness $A$ and (d) magnetocrystalline anisotropy $K_u$ as the out-of-plane electric field $\varepsilon$ sweeps from 0.5 V/Å to −0.5 V/Å (black points), and vice versa (red points). The blue right axes in (b)~(c) indicate the values of magnetic parameters with converted units. Inset in (b) presents the preferred chirality (indicated by the magenta arrows) of spin spiral for anticlockwise ($D > 0$) and clockwise ($D < 0$) DMI.

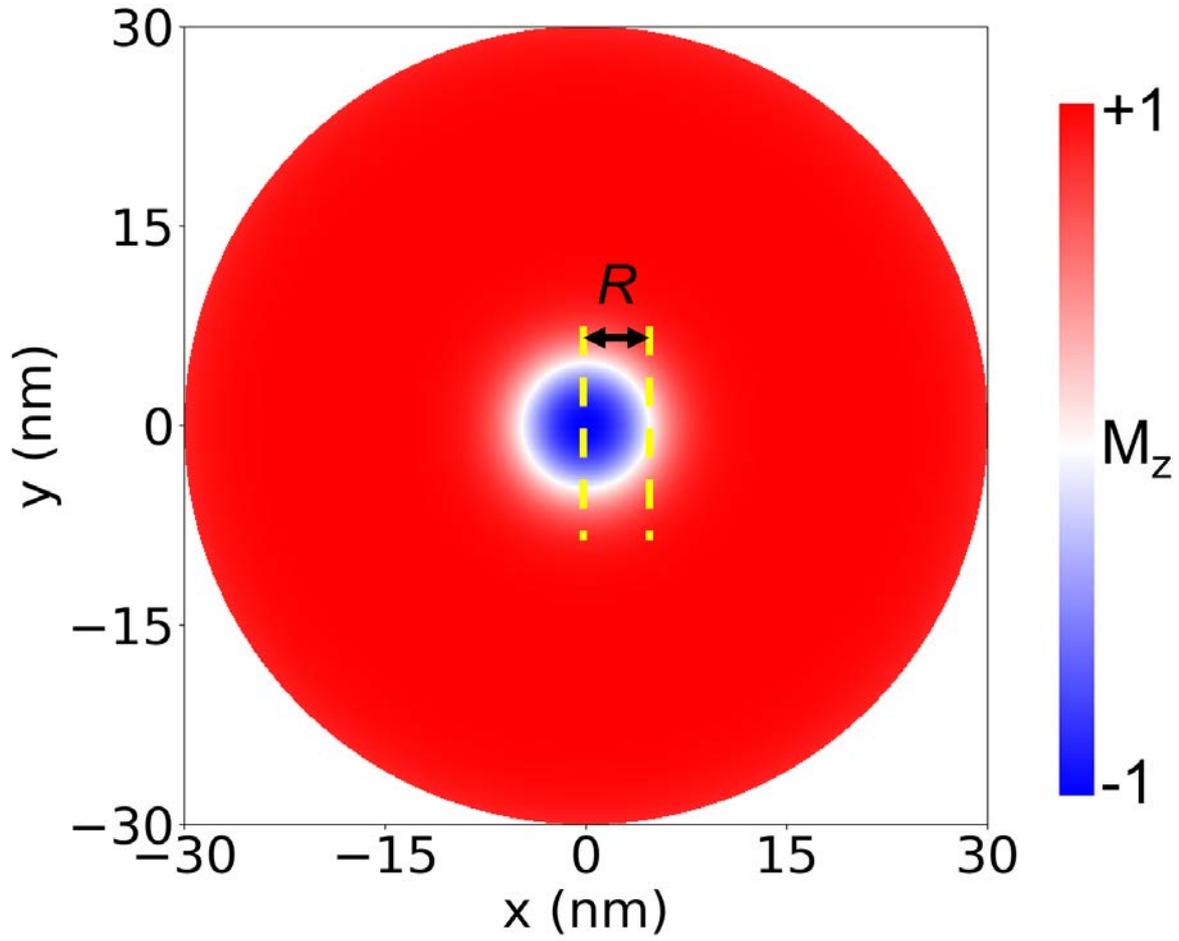

**Figure 5.** The relaxed skyrmionic magnetization configuration in a CrN nanodisk with radius of 30 nm in the absence of electric field. The out-of-plane component $m_z$ is indicated by the color map. The radius $R$ (defined as the radius of $m_z = 0$ contour) of relaxed skyrmion is 4.8 nm.